\documentclass{article}
\usepackage[T1]{fontenc} % add special characters (e.g., umlaute)
\usepackage[utf8]{inputenc} % set utf-8 as default input encoding
\usepackage{ismir,amsmath,cite,url}
\usepackage{graphicx}
\usepackage{color}

\usepackage{subcaption}
\usepackage{array}
\title{Towards Practical Automatic Piano Reduction using BERT with Semi-supervised Learning}

\multauthor
{Wan-ki Wong$^1$ \hspace{0.6em} Ka-ho To$^2$ \hspace{0.6em} Chuck-jee Chau$^2$
 \hspace{0.6em} Lucas Wong$^3$ \hspace{0.6em} Kevin Y. Yip$^2$ \hspace{0.6em} Irwin King$^2$}
{$^1$ Department of Information Engineering, The Chinese University of Hong Kong\\
 $^2$ Department of Computer Science and Engineering, The Chinese University of Hong Kong\\
 $^3$ Fazioli Artist\\
 \texttt{\small 1155124843@link.cuhk.edu.hk, 1155125547@link.cuhk.edu.hk, chuckjee@cse.cuhk.edu.hk,}\\
 \texttt{\small lucastwong@icloud.com, kevinyip@cse.cuhk.edu.hk, king@cse.cuhk.edu.hk}
}

\begin{document}

\maketitle
\begin{abstract}
In this study, we present a novel automatic piano reduction method with semi-supervised machine learning. Piano reduction is an important music transformation process, which helps musicians and composers as a musical sketch for performances and analysis. The automation of such is a highly challenging research problem but could bring huge conveniences as manually doing a piano reduction takes a lot of time and effort. While supervised machine learning is often a useful tool for learning input-output mappings, it is difficult to obtain a large quantity of labelled data. We aim to solve this problem by utilizing semi-supervised learning, so that the abundant available data in classical music can be leveraged to perform the task with little or no labelling effort. In this regard, we formulate a two-step approach of music simplification followed by harmonization. We further propose and implement two possible solutions making use of an existing machine learning framework---MidiBERT. We show that our solutions can output practical and realistic samples with an accurate reduction that needs only small adjustments in post-processing. Our study forms the groundwork for the use of semi-supervised learning in automatic piano reduction, where future researchers can take reference to produce more state-of-the-art results.
\end{abstract}
\section{Introduction}\label{sec:introduction}

Piano reduction is the process of transforming an orchestral or open score, with multiple instruments and/or voices, into a piano score that can be played with two hands. The task of automatic piano reduction is a lesser discussed, yet very important task in the field of Music Information Retrieval (MIR). Having a good piano reduction enables solo players to appreciate orchestral masterpieces in the world, extends repertoire of piano performances, as well as allowing more flexible arrangements for smaller ensembles to perform a piece. To composers, an easy and automatic piano reduction algorithm could bring huge convenience to their creative composition process, as they can easily analyze and revise their written melodies and harmonies by playing on the piano. We, as both computer scientists and music enthusiasts, are therefore very interested in solving this problem with automatic methods.

The major challenge of using machine learning methods for piano reduction lies in data labelling. There are currently no available datasets to obtain a large quantity of quality orchestral scores and their corresponding piano reduction scores. Manual labelling of data is extremely time consuming in piano reduction, and there is no model answer for a piano reduced score. It is hence not feasible to create a large, labelled training dataset. We therefore resort to semi-supervised training methods in our study.

In this study, we present and analyze our experiments with Bidirectional Encoder Representations from Transformers (BERT) to see how they can be modified and fine-tuned for the task of piano reduction. The objective is to take a multi-track MIDI orchestral score as input and produce a two-track MIDI piano score (one track for each hand) as an output. Our main contribution is the utilization of semi-supervised machine learning models, as we discover that there is large amount of openly available MIDI music data. On the other hand, paired orchestra and piano data are extremely limited and it would be extremely costly for us to manually obtain the necessary labels for supervised machine learning.

Our strategy of ``music simplification followed by harmonization'' adopts and combines multiple MIR techniques, such as melody extraction and music generation, together for the task of piano reduction. The resultant model is able to extract important notes, harmonize, and enrich the reduced music through learning from large quantity of data. Our solution also allows possibilities of adjustable difficulties and styles through further fine-tuning on the models.

%We have mainly formulated three different approaches, namely ``Melody extraction and Accompaniment Suggestion'', ``Direct conversion'' and ``Reduction then Generation''. Then, we proposed four different solutions corresponding to the three approaches, namely a rule-based method of building an accompaniment database and extracting melody with skyline algorithm (\texttt{DBM|), the use of CycleGAN~\cite{p4} for direct conversion from orchestra to piano score (\texttt{CG|), MidiBERT~\cite{p11} for deciding notes to be kept or discarded (\texttt{MB-NR|), and MidiBERT as a sequence-to-sequence model to turn an extracted melody and bass line of orchestra score into a full piano score (\texttt{MB-R2F|). The general workflow is presented in Figure \ref{fig:workflow}.\\

% Section 2 discusses the related works that have also worked on and proposed solutions for automatic piano reduction. Section 3 explains our data collection method and data sources. Section 4 provides a detailed explanation of our explored methods and implementation. Section 5 further explains the post-processing module that cleans up the model output and assign notes to two hands appropriately. Section 6 does a detailed evaluation on each solution. Finally, section 7 explains the difficulty encountered in the project and suggests future work for further improvement of models.

%
\section{Background}\label{sec:background}
%Several other studies have also attempted to generate a piano reduction score from an orchestra score using machine learning or algorithmic methods. Here we present some of the approaches and explain the relation of our work with theirs.

Most related work tackling the problem of piano reduction uses rule-based or statistical methods. Nakamura and Sagayama~\cite{p17} proposed an automatic piano reduction approach based on the Hidden Markov Model (HMM). They formulated the reduction task as an optimization problem, where the goal is to optimize the fidelity of the original score subject to the constraint of performance difficulty. Takamori \textit{et al.}~\cite{p18} proposed another rule-based method for pop music reduction, in which they treated the generation of right-hand and left-hand part of the piano score as two separate tasks. For the left-hand part, they constructed an accompaniment database from pop musical scores. For the right-hand part, they extracted the melody from vocal scores, which by default contains only the melody, then they added additional notes based on the chord in high accent locations. Due to the simplicity of Takamori \textit{et al.}’s method, we adapt and re-create it in the context of classical music, then treat it as a baseline model to compare with our PTM models. 

To our best knowledge, pre-trained models (PTMs) have never been implemented in the task of piano reduction before. PTMs like BERT~\cite{p15} are commonly used in natural language processing tasks which is proven to have achieved state-of-the-art results.

PTMs with the BERT architecture are first introduced in the MIR domain as in Chou \textit{et al.}’s MidiBERT model~\cite{p11} and Zeng \textit{et al.}’s MusicBERT model~\cite{p12}. Their models succeeded in utilizing large-scale musical data for pre-training and fine-tuning in a series of downstream tasks for symbolic music understanding. Their work, specifically the tokenization of musical data and the model architecture, has been very helpful to our study. We adopt the MidiBERT model as the base model and build on top of it for the piano reduction task. Our work also introduces the use-case of BERT models as a Seq2Seq model for music generation, which has not been discussed in previous works.

\section{Methodology}\label{sec:methodology}

\subsection{Data Sources}\label{subsec:data-sources}

%MIDI data  to leverage a larger amount of openly available data, we decided to use the MIDI format instead of MusicXML. MIDI is a format for representing symbolic musical data by a sequence of ``note on'' and ``note off'' events, also containing information such as instruments, tracks, tempo etc. While the MIDI format are not intended for representing a full musical score for musicians to play (rather a common protocol for electronic musical devices), we are able to gain a much larger corpus in this format, and the information contained would suffice for machine learning training.

We use MIDI symbolic music data of classical music in this study, summarized in Table~\ref{tab:datasource}. Piano files which has more than three tracks, and orchestra files which has less than four tracks are discarded (as they would cause scores in the two domains to be too similar). In our experiments, orchestra scores are merged to one single MIDI channel. 

\begin{table}[h]
 \begin{center}
 \begin{tabular}{p{2.5cm}p{5cm}}
Name  & Description\\
\hline
Live Orchestral Piano (``LOP'') database~\cite{p3} & Around 200 pairs of aligned orchestra scores and its corresponding piano arrangement.\\
\hline
GiantMIDI-Piano (``GiantMidi'') dataset~\cite{p5} & Around 10,000 piano scores, with mediocre quality due to limitations of the transcription method used by the author.\\
\hline
Symbolic Orchestral Database (``SOD'')~\cite{p3}   & Around 6,000 multi-instrument orchestral pieces.   Many different arrangements are covered, such as string quartets, SATB vocal scores, full orchestra etc.\\
\hline
Piano Midi~\cite{p6}                          & 200 piano scores with better quality, where scores are nicely transcribed with separation of right hand and left hand   in different tracks.\\
\hline
 \end{tabular}
\end{center}
 \caption{Various data sources in our project.}
 \label{tab:datasource}
\end{table}

\subsection{High-level Approach}
While designing our solutions, we follow a two-step approach of music simplification followed by harmonization.
The simplification step aims to choose a minimal subset of notes from the original orchestra score that will be kept in the piano reduction, while preserving the overall music fidelity. Then, the harmonization step adjusts the kept notes and adds additional notes to make the score complete and playable. A simpler rule-based method is also implemented for a baseline comparison.

Finally, we have a post-processing module to clean up the output scores and ensure their playability.% It will be discussed in length in Section~\ref{sec:postprocessing}.

%\subsection{Experiments}

%We aim to treat the extraction of melody and pairing of a reasonable accompaniment as two separate tasks. We have constructed a rule-based solution by referencing work of Takamori \textit{et al.}~\cite{p18} which involves approximating the orchestra score’s melody using the skyline algorithm, and constructing an accompaniment database to search for appropriate accompaniments. 

%The reduction part aims to choose a subset of notes from the original orchestra score that will be kept in the piano reduction, then the generation part adjusts the kept notes and add additional notes to make the score complete and playable. We have mainly explored MidiBERT in both reduction and generation tasks: for reduction, we tried to fine-tune MidiBERT to make a decision to keep or discard each note directly; for generation, we tried to turn MidiBERT into a sequence-to-sequence (Seq2Seq) model to generate a full sequence of piano score from a reduced score.

%In addition to the above methods, we also have a post-processing module to clean up the output scores and ensure their playability. It will be discussed in length in section 5.\\\\

\subsection{Experiments with the BERT Model}

\begin{figure*}[!hbt]
 \centerline{
 \includegraphics[width=1\textwidth]{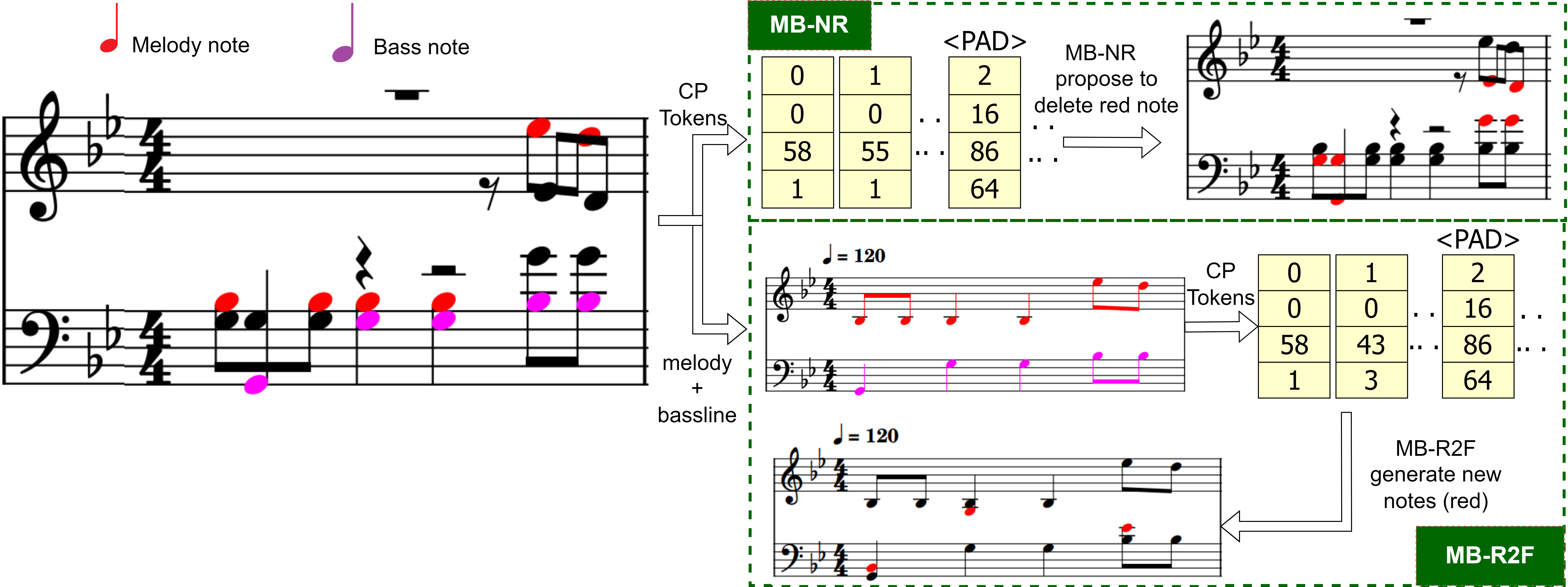}}
 \caption{Basic workflow of \texttt{MB-NR} and \texttt{MB-R2F}.}
 \label{fig:workflow}
\end{figure*}

%In this section, we try to overcome the difficulties of both the limited number of paired training samples and the limited computational power we have by using some available pre-trained models. In section 4.2, we have demonstrated how a pre-trained module can be integrated into the model and speed up convergence. We believe that we could also benefit from other pre-trained models that is trained with huge amount of musical data by other researchers.\\\\

We pre-train the MidiBERT model with classical music data to increase the relevance of the model with the genre. Then we explore various downstream tasks and turn it into a sequence-to-sequence (Seq2Seq) model, and propose two different solutions described in the following sections, illustrated in Figure~\ref{fig:workflow}.

%First, we fine-tuned MidiBERT on a new downstream task - note reduction, where the model decides on a subset of notes to be kept (\texttt{MB-NR|). It is a binary classification on every note in the orchestra score. Each note will either be removed from the score or be kept for the reduced piano score.

%Second, we stacked two MidiBERT models together, one as an encoder while the other acts as a decoder, to form a Seq2Seq model. The idea of turning a smaller BERT model (which is originally an encoder model only) to a larger seq2seq model comes from Chen \textit{et al.}'s~\cite{p25} study. The model is asked to generate additional notes over the given reduced score, obtained by only getting the melody and bass line from an orchestra score with skyline algorithm (\texttt{MB-R2F|). The exciting idea we bring here is that we can use self-supervised learning to train the model on large amount of piano data. We  need to extract the melody and bass line of a piano score as input, and ask the model to reconstruct the full score.\\\\

Input and output are tokenized using the Compound Word (CP) token representation from MidiBERT~\cite{p11}. Each note is represented as a token containing four dimensions as follows:

\begin{enumerate}
    \item \textit{New bar} (0 or 1): 1 means that the note is in the same bar as the previous token, 0 otherwise
    \item \textit{Position} (0--15): Number of sub-beats the note starts after the bar starts
    \item \textit{Pitch} (0--85): Relative pitch value of the note
    \item \textit{Duration} (0--63): Number of sub-beats
\end{enumerate}

Special tokens are used to facilitate the training process:
\begin{itemize}
    \item $\langle BOS\rangle$ indicates the beginning of the sequence
    \item $\langle EOS\rangle$ indicates the end of the sequence
    \item $\langle PAD\rangle$ prevents the model from applying attention to the token
    \item $\langle MASK\rangle$ masks the token (in pre-training) and the model should predict its original value
    \item \textit{$\langle ABS\rangle$ indicates that the current bar has no notes}
\end{itemize}

Here we propose the $\langle ABS\rangle$ token to solve the missing bar problem in CP representation. Since a new bar can only be indicated with a token while a token must correspond to a note in CP, if there is a bar with all rests for all instruments, then the whole bar will go unrepresented and become missing when reconstructing the score from CP tokens. The $\langle ABS\rangle$ token here represents an empty bar.

For each input MIDI file, every note is tokenized into a 4-tuple as defined above. The tokenized notes should be sorted by start time then by pitch.

\subsubsection{Pre-training}
% \begin{itemize}
% \setlength\itemsep{-0.5em}
%     \item Input token sequence:  \texttt{(batch\_size,512,4)}
%     \item Output token sequence: \texttt{(batch\_size,512,4)}
% \end{itemize}
In pre-training, we transform tokens into input and output sequences by concatenating them into sequences of length 512. The implementation of the pre-training task is similar to the approach used by the original authors of MidiBERT. We follow the mask language modelling (MLM) approach to randomly mask tokens and ask the BERT model to predict those tokens in the pre-training stage. The only difference in our pre-training is that the GiantMidi and SOD datasets are used to create more relevance to classical music.

\begin{figure}[!htb]
 \centerline{
 \includegraphics[width=1.1\columnwidth]{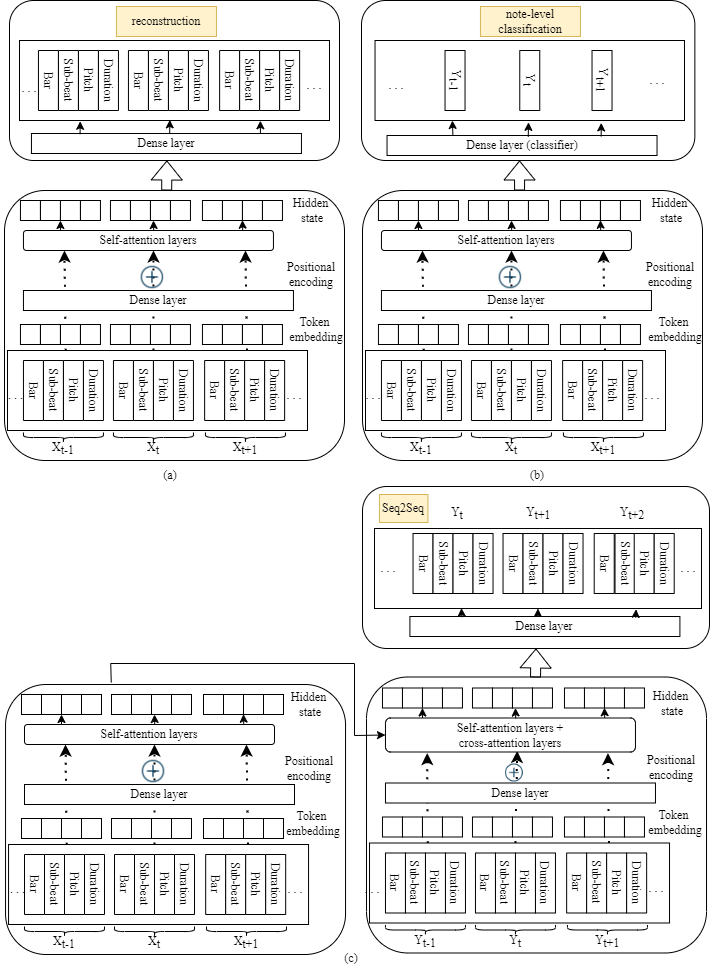}}
    \caption{Architecture of MidiBERT~\cite{p11}, (a): Pre-training, (b): \texttt{MB-NR}, (c): \texttt{MB-R2F}.}
    \label{fig:architecture}
\end{figure}

The structure of MidiBERT in our tasks can be seen in Figure~\ref{fig:architecture}. For in-depth details, we refer readers to the original MidiBERT paper~\cite{p11}.

\begin{figure*}[!htb]
\centering
\begin{subfigure}{.33\textwidth}
  \centering
  \includegraphics{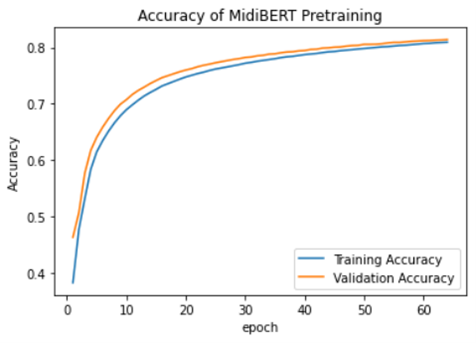}
  \caption{Pre-training Accuracy}
\end{subfigure}%
\begin{subfigure}{.33\textwidth}
  \centering
  \includegraphics{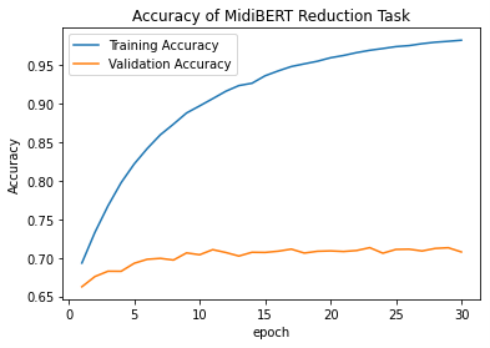}
  \caption{\texttt{MB-NR} accuracy}
\end{subfigure}
\begin{subfigure}{.33\textwidth}
  \centering
  \includegraphics{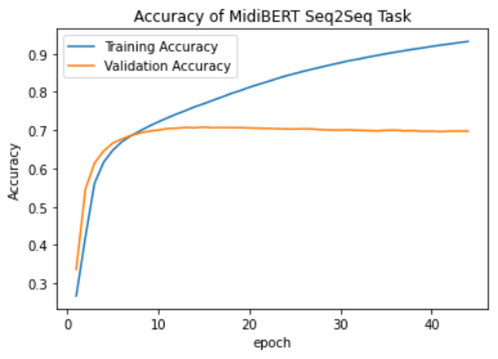}
  \caption{\texttt{MB-R2F} accuracy}
\end{subfigure}
 \caption{Accuracy graphs for model training.}
 \label{fig:accuracy}
\end{figure*}

\subsubsection{Solution 1: Note reduction ``\texttt{MB-NR}''}
% \begin{itemize}\setlength\itemsep{-0.5em}
%     \item Input token sequence:  \texttt{(batch\_size,512,4)}
%     \item Output sequence : \texttt{(batch\_size,512)}
% \end{itemize}

For note reduction as a downstream task, the input sequences are prepared similar to the format of pre-training and solely from orchestra data in the LOP database. The output has the same length as the input, where each output (0/1) indicates whether to keep or discard the corresponding note from the input sequence.

To generate ground-truth labels for this task, we utilize the aligned and paired piano samples from the LOP dataset. Each orchestra note is marked as to keep only if it has intersection with any note in the corresponding piano score with the same pitch.

The structure of MidiBERT on the note reduction task is almost identical to the pre-training task, except that the final dense layers are replaced with a single classifier to convert the hidden dimension into a binary output (0/1).

The output is compared to the ground truth using a binary-cross-entropy loss. The gradients are propagated with an Adam optimizer with weight decay.

\subsubsection{Solution 2: Seq2Seq generation ``\texttt{MB-R2F}''}
% \begin{itemize}\setlength\itemsep{-0.5em}
%     \item Input token sequence:  \texttt{(batch\_size,512,4)}
%     \item Output token sequence: \texttt{(batch\_size,512,4)}
% \end{itemize}

In this solution to turn a reduced score into a full piano score, we use the skyline algorithm\cite{p20} to extract the bass line and melody line from the piano score. The input sequences are prepared from the extracted notes in a way similar to Solution 1, but each input sequence is restricted to only contain at most 90 non-$\langle PAD\rangle$ tokens (the remaining length all contains $\langle PAD\rangle$ tokens). This is because we need to have a 512-length corresponding output of full piano score that could potentially contain many more notes than the input. For the ground-truth output sequence, we convert the corresponding section of the full score into tokens, then add the $\langle BOS\rangle$ token in front and $\langle EOS\rangle$ token at the back. We discard the whole input-output pair if the output sequence length exceeds 512. 

Here we chain two identical MidiBERT together to build a Seq2Seq model.
The Seq2Seq model consists of three parts. First, we use one MidiBERT as the encoder. Then we use another MidiBERT as the decoder by adding cross-attention layers to it, so that it can attend to the hidden vectors of the encoder. Finally, dense layers are added on top of the decoder output, which are responsible for returning the decoder's hidden states to CP tokens.

During training, the input and output sequences are provided to the encoder and decoder respectively. Huggingface~\cite{p26} (the library that implements the BERT models) has implemented the causal mask for us to avoid the model gaining ``future'' information on what it needs to predict. The model is expected to produce a left-shifted-by-1 version of the output sequence, such that the decoder can learn to always produce the next token given the previous sequence input. We train the model this way using a cross entropy loss which is aggregated for all four dimensions of the token, and we monitor its accuracy across the four dimensions.

Since the input and output sequences could be of different length, we rely on the special token $\langle EOS\rangle$ to determine where the output sequence ends. An iterative prediction is used during inference. Every input sample of shape \texttt{(1,512,4)} goes through multiple prediction rounds. First, the input sequence is fed into the encoder to obtain the hidden representations. At the beginning of inference, the output sequence only contains a single $\langle BOS\rangle$ token. We feed the encoder’s hidden vector and the latest output sequence (padded to 512 length) into the decoder and obtain the next token in the output. The predicted token is appended to the output sequence and fed to the decoder again to get the subsequent token. The loop ends when the $\langle EOS\rangle$ token is predicted or when the length of the output sequence exceeds 512.

\subsubsection{Results and Analysis}
\label{sec:results}

Thanks to the pre-trained model checkpoints, we can achieve good accuracy in both \texttt{MB-NR} and \texttt{MB-R2F} within a very short training period. Figure~\ref{fig:accuracy} shows the accuracy graph of all the tasks. It is observed that for the pre-training task, the validation accuracy is higher than the training accuracy, which might be due to the model being trained on musical data already. It also supports the claim of BERT models that by using a pre-trained model, we can try to train a working model based on our needs with a minimal amount of data. Readers can access audio and score samples at this link\footnote{URL is removed for anonymity and samples can be found as supplementary materials.}.

\begin{figure}[h]
 \centerline{
 \includegraphics[width=1\columnwidth]{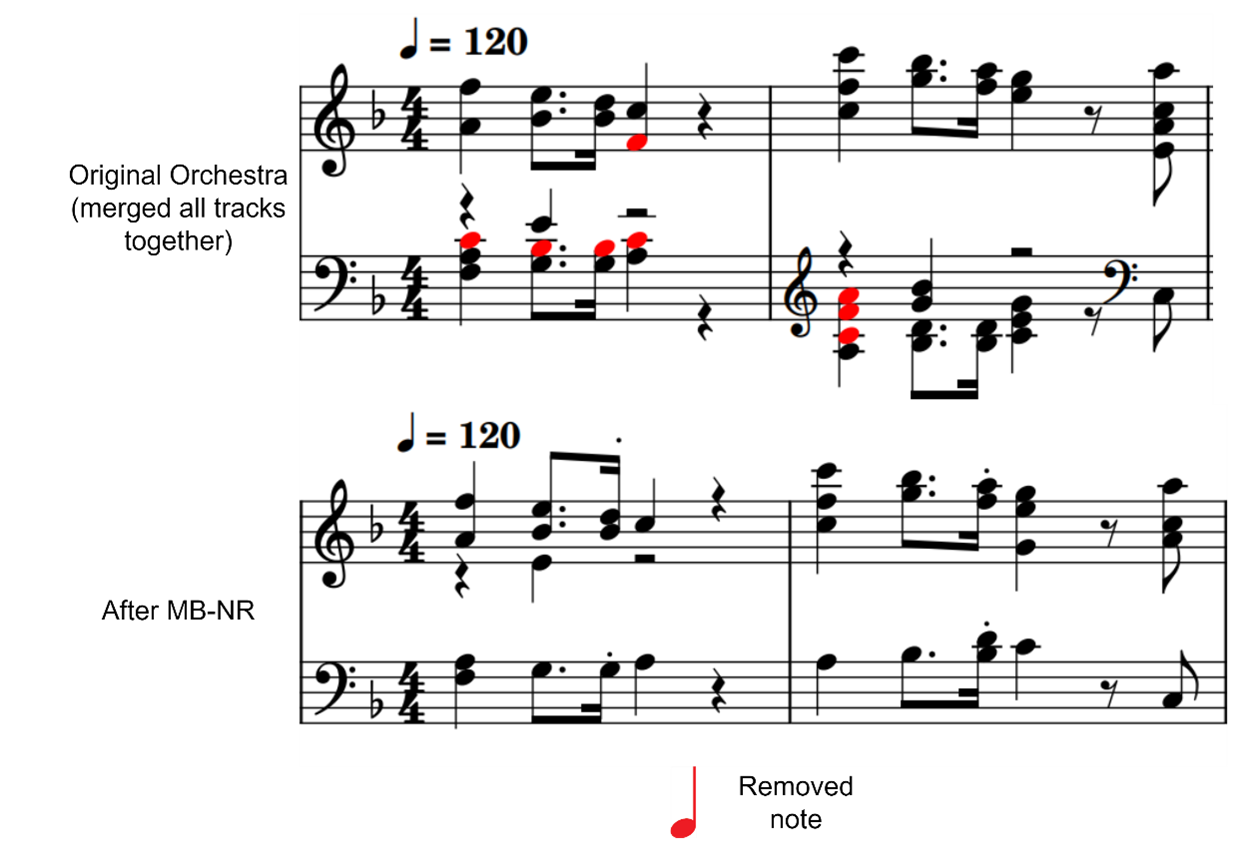}}
 \caption{Analysis of \texttt{MB-NR} result of a good example.}
 \label{fig:MBNRgood}
\end{figure}

\paragraph*{Solution 1: \texttt{MB-NR}} 
%Two examples of \texttt{MB-NR} reduction are shown in the Appendix. 
In Figure~\ref{fig:MBNRgood}, we can see a good reduction sample which is able to conserve all the melody and bass notes. The removed notes are the intermediate chord notes that has minimal impact to the listener even if they are removed.
Yet, \texttt{MB-NR} may sometimes misunderstand the music content, resulting in a faulty reduction %shown in Figure~\ref{fig:MBNRbad} in Appendix. 
where we notice that it accidentally removes the melody note and the bass note. %Consequently, it causes an ambiguity in the chord highlighted in red, affecting the chord progression of the song and making it less tonally similar from its input. Since it does not introduce dissonance to the music and the reduction error is mainly due to the chord note choice, we believe that the error mentioned here could be mitigated if we provide the chord information as a future work.

\paragraph*{Solution 2: \texttt{MB-R2F}}
%One example is shown in Figure~\ref{fig:MBR2F} in Appendix to demonstrate the result of the \texttt{MB-R2F} model. 
One thing worth noticing is that all the notes generated by the \texttt{MB-R2F} solution is solely based on the notes obtained from the skyline algorithm. If we only compare the \texttt{MB-R2F} output with the orchestra input, the model is able to reduce non-melody and non-bass notes from the original score. It is also observed that all the generated notes are aligned, of harmony with the original chords, and is never placed beyond the range of the original input. Hence we regard the generation output to be quite successful.

\subsection{Baseline with rule-based solution ``\texttt{DBM}''}
\label{sec:baseline}
We construct a rule-based solution by referencing Takamori \textit{et al.}~\cite{p18} on approximating the orchestra score’s melody using the skyline algorithm, and build an accompaniment database to search for appropriate accompaniment styles in classical music.

Our baseline solution works as follows. For the right-hand part, we mainly refer to Ozcan, Isikhan and Alpkocak's approach ~\cite{p19} to select appropriate melody channels, combined with Chai's improved skyline algorithm~\cite{p20} to accurately extract the melody from a multi-track MIDI orchestra score.
For the left-hand part, we utilize the ``Piano Midi'' dataset to extract possible accompaniment styles in classical piano music instead of pop music. Then, we follow Takamori \textit{et al.}'s method to match appropriate entries from the database to a particular bar of the piano reduced score.

This method basically ensures the playability of the piece and shows good resemblance to the original score. Details are not covered here due to space limitations.

%%%%%%%%%%%%%%%%%%%%
\subsection{Post-processing}
\label{sec:postprocessing}
The post-processing module is responsible for tidying up the output piano score. As such, the design goal is to tackle the reduction errors to increase their readability and playability. It runs iteratively until the score cannot be further simplified or intentionally terminated by the user.

\paragraph*{Clustering} 
To mimic the two hands of a pianist, we perform clustering on the notes beat by beat to group them into either for the left hand or right hand. We first collect all the notes inside each beat, and then find two centres of distribution by Kernel Density Estimation based on their pitches. The notes are then clustered into two groups based on their nearest centre. We start from a large kernel that usually groups all the notes into one cluster. Then, we decrease the kernel size gradually until it identifies two peaks, which represent the two centres. The centre with a lower pitch value becomes the left hand's centre.

\paragraph*{Simplifying} 
We limit the maximum note duration to four beats, and notes longer will be trimmed. It is because the sound of a note fades out exponentially on a piano. Trimming or removing long notes can increase the playability of the score as the player does not need to hold a long note, and they can move their hands around with fewer constraints. We then delete doubled notes when more than four notes are played simultaneously on each hand. We consider only the notes that starts at the same tick and shares the same pitch class as a pair of doubled notes. 

\paragraph*{Transposing} 
After finding a centre of left and right-hand, we assign each note to its nearest centre. If the note is too far away from the corresponding centre, it is transposed an octave towards the centre, bounded by the original highest/lowest pitches. To avoid breaking the melody and bass pattern, we do not allow notes on the right hand to transpose downward and the note on the left hand to transpose upward. 
After transposition, a gap may appear which hurts playability and musicality if the two hands’ centres are too far apart. All left-hand notes are transposed upward for an octave if the single linkage between two groups is greater than 18 semitones, which is larger than an octave. %($min\_pitch(right\_hand) - max\_pitch(left\_hand) > 18$)

%%%%%%%%%%%%%%%%%%%%%%%%
\section{Evaluation}
\subsection{Objective Metrics}

\subsubsection{Tonal Similarity}
The tonal similarity metric aims to calculate the similarity between the reduced piano score and the original orchestra score by looking at the pitch-class histogram of the two scores with a sliding window of two beats. The use of pitch-class histograms to evaluate tonal distances has been proposed by Tenkanen~\cite{p30}, which could effectively help determine similarity between two pieces.
This metric must range between 0 and 1, from the most dissimilar to the most similar.

We take 77 orchestral pieces at random, pass them through our solutions and compute the input/output tonal similarity with this metric. The performance of our solutions are summarized in Table~\ref{tab:similarity}.\\

\begin{table}[!htb]
 \begin{center}
 \begin{tabular}{lccc}
           & \texttt{MB-NR} & \texttt{MB-R2F} &\texttt{DBM}\\
           \hline
Mean & $0.39$  & $0.58$  & $\mathbf{0.74}$  \\
s.d. & $0.09$  & $0.15$  & $0.07$ \\
Max  & $0.88$  & $0.93$  & $0.91$     \\
Min  & $0.18$  & $0.27$ & $0.51$  \\
\hline
 \end{tabular}
\end{center}
 \caption{Tonal similarity of the solutions, 0: least similar, 1: most similar.}
 \label{tab:similarity}
\end{table}

The \texttt{MB-NR} solution produces the least tonally accurate reductions. This might be because the model only selects notes to be dropped from the original score but does not adjust the total duration of other notes or add additional notes, causing the pitch histogram to be almost always different from the original score. This shows a weakness of the model, which is the inability to maintain the relative proportion of different pitch classes.

\begin{table*}
 \begin{center}
 \begin{tabular}{l|cc|cc|cc}
           & \multicolumn{2}{>{\centering\arraybackslash}p{.2\linewidth}}{\tt MB-NR}   & \multicolumn{2}{>{\centering\arraybackslash}p{.2\linewidth}}{\tt MB-R2F}& \multicolumn{2}{>{\centering\arraybackslash}p{.2\linewidth}}{\tt DBM} \\
Test case & 1            & 2        & 1          & 2         & 1            & 2     \\
           \hline
Accuracy  & $0.361$        & $0.443$        & $0.738$      & $0.541$        & $0.426$     & $0.607$ \\
s.d.      & $0.484$        & $0.501$        & $0.444$      & $0.502$     & $0.499$      & $0.493$ \\
t-value   & $\mathbf{-2.229}$       & $\mathbf{-0.887}$       & $4.151$      & $\mathbf{0.632}$        & $\mathbf{-1.146}$    & $1.676$ \\
\hline
Rejected? & Not rejected & Not rejected & Rejected   & Not rejected & Not rejected & Rejected  \\
\hline
 \end{tabular}
\end{center}
 \caption{Results of the discrimination test.}
 \label{tab:discrimination}
\end{table*}

\subsection{Subjective Metrics}
The subjective metrics are developed for a more wholesome evaluation on model performance, including the reduction quality and playability. We include subjective metrics as piano reduction is after all a very subjective task and there is no definitive answer on a ``best'' reduction.

\subsubsection{Discrimination Test}

We adopt the method of Pearce and Wiggins~\cite{p29} to evaluate our models by generating six pieces of system-arranged piano reductions (two pieces for each of our proposed solutions). We pair them with six pieces of human-arranged piano pieces taken from the LOP dataset and shuffle their order randomly. We then ask subjects (both musically trained and not trained) to classify whether each piece is arranged by human or computer.

For each sample generated by our solutions, we calculate its mean accuracy as \% of subjects answering ``computer''. We formulate the null hypothesis as that the sample has successfully tricked people into believing that it was generated by a human. We consider our solution to pass the discrimination test and has the most realistic output if the null hypothesis is not rejected for both samples corresponding to the model.

We perform statistical testing on the above hypothesis with the one-tailed student’s t-test and a confidence interval of 0.05. As we have collected 61 samples, the degree of freedom for the testing is 60 and the rejection threshold is set to 1.67 for all the test cases. The results are summarized in Table~\ref{tab:discrimination}.

We can observe that our \texttt{MB-NR} solution successfully passes the discrimination test, where both samples have achieved a very low accuracy. This shows that \texttt{MB-NR} generates the most realistic samples amongst all our solutions. For the two remaining models, one test case is rejected while the other is not rejected, showing that the model performance still fluctuates among different samples.

\subsubsection{Professional Evaluation}

We also sent out surveys to subjects who are professional pianists to grade the piano scores produced by our three solutions. The criteria we use are designed by Nakamura and Sagayama~\cite{p17} for the evaluation of their piano reduction model, and are listed below:
\begin{itemize}
\setlength\itemsep{-0.4em}
    \item \textit{Musical Fidelity}: How faithful is it to the original music? 1: Not faithful at all, 5: Very faithful
    \item \textit{Difficulty}: How difficult is it to play the score on the piano with two hands. 1: Very easy, 5: Very difficult
    \item \textit{Naturalness}: How natural is the reduction score as a piano score? 1: Very unnatural, 5: Very natural
    \item \textit{Reduction Quality}: How much adjustment is needed for the score to be an acceptable piano reduction? 1: Very high degree of adjustments is needed, 5: No adjustments are needed at all
\end{itemize}
We have successfully collected 10 responses and calculated the mean score for each question. The results of the survey are summarized in Table~\ref{tab:profevaluation}.

\begin{table}[h]
 \begin{center}
 \begin{tabular}{lccc}
           & \texttt{MB-NR} & \texttt{MB-R2F} &\texttt{DBM}\\
           \hline
Musical Fidelity  & $\mathbf{3.8}$     & $3.4$ & $3.1$    \\
Difficulty        & $2.9$   & $2.5$ & $\mathbf{3.2}$    \\
Naturalness       & $\mathbf{3.6}$     & $3.4$     & $2.3$  \\
Reduction Quality & $\mathbf{3.6}$     & $3.4$  & $2.8$  \\
\hline
 \end{tabular}
\end{center}
 \caption{Results of professional evaluation.}
 \label{tab:profevaluation}
\end{table}

The results show that \texttt{MB-NR} has the best performance when viewed by professional pianists. It has the highest score in musical fidelity (showing that it reconstructs the original music well), naturalness (showing that it is very similar to a human-arranged piano score), and reduction quality (showing that the number of manual adjustments needed after model output is small). It has  also an average difficulty, showing that it is not too hard and not too easy to play. The \texttt{DBM} and \texttt{MB-R2F} solutions have slightly worse performance.

We invited musicians and professional to provide comments to each reduction score as an informal qualitative assessment. For both \texttt{MB-NR} and \texttt{MB-R2F}, there were comments mentioning the reductions are not bad and sounds reasonable. Audio and score samples can be accessed via the link in Section~\ref{sec:results}.

%\subsection{Evaluation Summary}
%To summarize the different evaluation attempts made in this section, we find that \texttt{MB-NR} performs the best amongst all solutions. It is the best-evaluated solution by professional pianists, and it fully passes our discrimination test. With \texttt{MB-NR|, the piano score must be a subset of the orchestra score with no additional notes, hence the average dissonance would be acceptable, and the lower tonal similarity can be compensated (i.e. people will still feel it as the same song despite the difference in pitch class proportions as there are no additional notes that do not appear in the original orchestra).

%The \texttt{MB-R2F} solution also have acceptable performance. The scores given by professional pianists on its overall quality, naturalness and musical fidelity are good (all over an average of 3).

%Both \texttt{DBM} and \texttt{MB-R2F} partially pass the discrimination test and have a reasonable result in average dissonance as well as tonal similarity. Their output is valuable as well since they could potentially provide more variations than the \texttt{MB-NR} model while having a reasonable and playable score. We leave the fine-tuning and further adjustments of the two solutions as future work.

%%%%%%%%%%%%%%%%%%%%
\section{Conclusion}
We have proposed two solutions for automatic piano reduction. The \texttt{MB-NR} solution is a simpler solution that mainly focuses on note reduction, which is shown to be a feasible approach as professional pianists favour this method the most. The \texttt{MB-R2F} solution generates new notes from the melody and bass line, which we have shown to be performing pretty well, as most notes generated are aligned and in harmony. From our overall evaluation results, the \texttt{MB-NR} method achieves the most acceptable and realistic performance while both the \texttt{MB-NR} and \texttt{MB-R2F} solution performs holistically better than our baseline solution \texttt{DBM}. This shows their promise in becoming a realistic model.

To summarize our contributions, we pioneer the exploration of semi-supervised machine learning methods for the task of piano reduction. We propose innovative methods like the \texttt{MB-R2F} approach for music harmonization in a piano score. We look forward to extensions by future researchers, such as to introduce variations in the reduction output and adjustable difficulty by fine-tuning our models. 

\section{Source Code}
All source code used in this experiment can be found in the following repository: \url{https://github.com/tpmmthomas/piano-reduction-with-midibert}.

% For bibtex users:
\bibliography{ref}

@article{p3,
  title={A database linking piano and orchestral MIDI scores with application to automatic projective orchestration},
  author={Crestel, L{\'e}opold and Esling, Philippe and Heng, Lena and McAdams, Stephen},
  journal={arXiv preprint arXiv:1810.08611},
  year={2018}
}

@article{p5,
  title={Giantmidi-piano: A large-scale midi dataset for classical piano music},
  author={Kong, Qiuqiang and Li, Bochen and Chen, Jitong and Wang, Yuxuan},
  journal={arXiv preprint arXiv:2010.07061},
  year={2020}
}

@misc{p6, title={PianoMidi}, howpublished={\url{http://www.piano-midi.de/midicoll.html}}, journal={Classical Piano Midi Page - Main Page}, author={Krueger, Bernd}}

@article{p11,
  title={MidiBERT-Piano: Large-scale Pre-training for Symbolic Music Understanding},
  author={Chou, Yi-Hui and Chen, I and Chang, Chin-Jui and Ching, Joann and Yang, Yi-Hsuan},
  journal={arXiv preprint arXiv:2107.05223},
  year={2021}
}

@article{p12,
  title={Musicbert: Symbolic music understanding with large-scale pre-training},
  author={Zeng, Mingliang and Tan, Xu and Wang, Rui and Ju, Zeqian and Qin, Tao and Liu, Tie-Yan},
  journal={arXiv preprint arXiv:2106.05630},
  year={2021}
}

@article{p15,
  title={Bert: Pre-training of deep bidirectional transformers for language understanding},
  author={Devlin, Jacob and Chang, Ming-Wei and Lee, Kenton and Toutanova, Kristina},
  journal={arXiv preprint arXiv:1810.04805},
  year={2018}
}

@inproceedings{p17,
  title={Automatic piano reduction from ensemble scores based on merged-output hidden markov model},
  author={Nakamura, Eita and Sagayama, Shigeki},
  booktitle={Proc. 41st Int. Comp. Music Conf. (ICMC)},
  year={2015}
}

@inproceedings{p18,
  title={Automatic arranging musical score for piano using important musical elements},
  author={Takamori, Hirofumi and Sato, Haruki and Nakatsuka, Takayuki and Morishima, Shigeo},
  booktitle={Proc. 14th Sound and Music Computing Conf., Aalto, Finland},
  pages={35--41},
  year={2017}
}

@inproceedings{p19,
  title={Melody extraction on MIDI music files},
  author={Ozcan, Giyasettin and Isikhan, Cihan and Alpkocak, Adil},
  booktitle={7th IEEE Int. Symp. Multimedia (ISM'05)},
  pages={8--pp},
  year={2005},
  organization={Ieee}
}

@inproceedings{p20,
  title={Melody retrieval on the web},
  author={Chai, Wei and Vercoe, Barry},
  booktitle={Multimedia Computing and Networking 2002},
  volume={4673},
  pages={226--241},
  year={2001},
  organization={Int. Soc. for Optics and Photonics}
}

@inproceedings{p26,
  title={Transformers: State-of-the-art natural language processing},
  author={Wolf, Thomas and Debut, Lysandre and Sanh, Victor and Chaumond, Julien and Delangue, Clement and Moi, Anthony and Cistac, Pierric and Rault, Tim and Louf, R{\'e}mi and Funtowicz, Morgan and others},
  booktitle={Proc. 2020 Conf. Empirical Methods in Natural Language Process.: System Demonstrations},
  pages={38--45},
  year={2020}
}

@inproceedings{p29,
  title={Towards a framework for the evaluation of machine compositions},
  author={Pearce, Marcus and Wiggins, Geraint and others},
  booktitle={Proc. AISB’01 Symp. Artificial Intell. and Creativity in the Arts and Sci.},
  pages={22--32},
  year={2001},
  organization={Citeseer}
}

@inproceedings{p30,
  title={Evaluating tonal distances between pitch-class sets and predicting their tonal centres by computational models},
  author={Tenkanen, Atte},
  booktitle={Int. Conf. Math. and Computation in Music},
  pages={245--257},
  year={2009},
  organization={Springer}
}

% For non bibtex users:
%\begin{thebibliography}{citations}
% \bibitem{Author:17}
% E.~Author and B.~Authour, ``The title of the conference paper,'' in {\em Proc.
% of the Int. Society for Music Information Retrieval Conf.}, (Suzhou, China),
% pp.~111--117, 2017.
%
% \bibitem{Someone:10}
% A.~Someone, B.~Someone, and C.~Someone, ``The title of the journal paper,''
%  {\em Journal of New Music Research}, vol.~A, pp.~111--222, September 2010.
%
% \bibitem{Person:20}
% O.~Person, {\em Title of the Book}.
% \newblock Montr\'{e}al, Canada: McGill-Queen's University Press, 2021.
%
% \bibitem{Person:09}
% F.~Person and S.~Person, ``Title of a chapter this book,'' in {\em A Book
% Containing Delightful Chapters} (A.~G. Editor, ed.), pp.~58--102, Tokyo,
% Japan: The Publisher, 2009.
%
%
%\end{thebibliography}

%\newpage
%\appendix

%\section{Samples of piano reduction results}

%\begin{figure}[h]
% \centerline{
% \includegraphics[width=1.2\columnwidth]{img/pic27.png}}
% \caption{Analysis of \texttt{MB-NR} result of a good example.}
% \label{fig:MBNRgood}
%\end{figure}

%\begin{figure}[h]
% \centerline{
% \includegraphics[width=1.2\columnwidth]{img/pic28.png}}
% \caption{Analysis of \texttt{MB-NR} result of a bad example.}
% \label{fig:MBNRbad}
%\end{figure}

%\begin{figure*}
% \centerline{
% \includegraphics[width=1.1\textwidth]{img/pic29.png}}
% \caption{Analysis of \texttt{MB-R2F} result.}
% \label{fig:MBR2F}
%\end{figure*}

\end{document}